\documentclass[conference]{IEEEtran}
\IEEEoverridecommandlockouts
\usepackage{cite}
\usepackage{amsmath,amssymb,amsfonts}
\usepackage{graphicx}
\usepackage{textcomp}
\usepackage{enumitem}
\usepackage{xcolor}
\usepackage{algorithmic}
\usepackage{algorithm}
\usepackage{xcolor}
\usepackage{multirow}
\usepackage{caption}
\usepackage{subfigure}
\usepackage{subcaption}
\usepackage{comment}

\def\BibTeX{{\rm B\kern-.05em{\sc i\kern-.025em b}\kern-.08em
    T\kern-.1667em\lower.7ex\hbox{E}\kern-.125emX}}
\begin{document}

\title{FLARE: A Wireless Side-Channel Fingerprinting Attack on Federated Learning}
\author{\IEEEauthorblockN{Md Nahid Hasan Shuvo\IEEEauthorrefmark{1}, Moinul Hossain\IEEEauthorrefmark{1}, Anik Mallik\IEEEauthorrefmark{2}, Jeffrey Twigg\IEEEauthorrefmark{3}, and Fikadu Dagefu\IEEEauthorrefmark{3}} \IEEEauthorblockA{\IEEEauthorrefmark{1}George Mason University, Fairfax, VA, USA} \IEEEauthorblockA{\IEEEauthorrefmark{2}Towson University, Towson, MD 21252, USA} \IEEEauthorblockA{\IEEEauthorrefmark{3}DEVCOM Army Research Laboratory, Adelphi, MD 20783, USA} \IEEEauthorblockA{ Email: mshuvo@gmu.edu, mhossa5@gmu.edu, amallik@towson.edu,\\ jeffrey.n.twigg.civ@army.mil, fikadu.t.dagefu.civ@army.mil}}
\maketitle

\begin{abstract}
    Federated Learning (FL) enables collaborative model training across distributed devices while safeguarding data and user privacy. However, FL remains susceptible to privacy threats that can compromise data via direct means. That said, indirectly compromising the confidentiality of the FL model architecture (e.g., a convolutional neural network (CNN) or a recurrent neural network (RNN)) on a client device by an outsider remains unexplored. If leaked, this information can enable next-level attacks tailored to the architecture. This paper proposes a novel side-channel fingerprinting attack, leveraging flow-level and packet-level statistics of encrypted wireless traffic from an FL client to infer its deep learning model architecture. We name it \textbf{FLARE}, a fingerprinting framework based on FL Architecture REconnaissance. Evaluation across various CNN and RNN variants—including pre-trained and custom models trained over IEEE 802.11 Wi-Fi—shows that FLARE achieves over 98\% F1-score in closed-world and up to 91\% in open-world scenarios. These results reveal that CNN and RNN models leak distinguishable traffic patterns, enabling architecture fingerprinting even under realistic FL settings with hardware, software, and data heterogeneity. To our knowledge, this is the first work to fingerprint FL model architectures by sniffing encrypted wireless traffic, exposing a critical side-channel vulnerability in current FL systems.

\end{abstract}

\vspace{-0.1in}


\section{Introduction}\vspace{-0.05in}
Federated learning (FL) enables multiple devices to collaboratively train deep learning models without exchanging raw data~\cite{mcmahan2017communication}. By keeping data locally stored on devices and only sharing encrypted model parameters, FL enhances user privacy in sensitive domains such as the Internet of Things (IoT), autonomous vehicles, and healthcare~\cite{kairouz2021advances}. Despite these privacy protections, FL remains vulnerable to privacy threats. Previous studies investigated direct attacks, such as data poisoning, membership inference, or model inversion, typically assuming the adversary already knows the underlying model architecture~\cite{kaushal2025securing,zhang2024survey}. However, little attention has been given to indirect privacy leaks through network side-channel analysis, particularly in wireless communication scenarios.

Network traffic analysis has long been leveraged by adversaries to infer sensitive information despite encryption. Prominent prior work includes website fingerprinting attacks on Tor or HTTPS traffic \cite{acar2020peek,sirinam2019triplet}. In the Web domain, as shown in \cite{wang2020high}, researchers demonstrate that even encrypted web traffic can be analyzed to identify visited applications and browsing activities. Device fingerprinting is another related area, where several studies have exposed how browser and system characteristics can uniquely identify IoT devices by analyzing their network traffic patterns~\cite{chowdhury2022survey}. Recent research into mobile app fingerprinting further shows that wireless traffic metadata can identify individual mobile application usage~\cite{li2024robust}. These works illustrate that metadata—packet sizes or timing—often contains distinguishing patterns that can undermine privacy.

\textbf{Motivation:} This research investigates the vulnerability of FL systems to similar privacy risks by analyzing encrypted network traffic. Given that different model architectures (e.g., CNNs and RNNs) comprise distinct computational graphs, parameter sizes, and training behaviors, their traffic volume, burst patterns, or timing characteristics during training could be inherently different. Such architectural fingerprinting attacks can further enable a range of downstream attacks, e.g., adversarial examples tailored to the detected model family, model inversion techniques targeting sequential data, or exposing known implementation vulnerabilities. Moreover, when combined with side-channel identifiers, e.g., MAC addresses, model fingerprinting can compromise user anonymity by mapping specific clients to model types. In Section \ref{sec:evaluation}, we show how fingerprinting can help an adversary throttle network throughput in an informed way to increase training convergence time. Given the widespread deployment of FL in privacy-sensitive domains, assessing this novel attack surface from an adversary's perspective is critical to understanding this vulnerability and securing future FL-enabled applications.

\textbf{Challenges:} Fingerprinting model architectures in wireless FL settings, however, presents several key challenges that must be addressed to ensure practical effectiveness.
\textbf{C1: Infrastructural Heterogeneity}—In real deployments, FL network traffic characteristics are influenced by clients' hardware resources, operating systems, and wireless conditions. These variations can introduce noise and timing discrepancies that may obscure architecture-specific patterns. A robust fingerprinting system must tolerate such platform-induced variability.
\textbf{C2: Model and Data Heterogeneity}—Traffic patterns are not determined solely by the model architecture but are also shaped by the dataset type, task complexity, and the federated learning pipeline itself. For instance, CNN and RNN models may exhibit overlapping traffic behaviors, and this effect is compounded by diverse datasets (e.g., STL10 vs. UCI HAR) and different aggregation strategies (e.g., FedAvg vs. FedProx). This makes isolating architecture-specific information more difficult.
\textbf{C3: Partial Observation}—A passive adversary may begin monitoring the FL process at any point in time, making it infeasible to capture a complete training session. The fingerprinting framework must therefore remain effective even with short or incomplete observations.

\textbf{Contributions:} In this work, we address these challenges by considering a wide range of neural architectures, including popular Deep CNN variants (e.g., ResNet18, MobileNetV2, DenseNet) and RNN variants (e.g., bidirectional LSTM, GRU, and a novel hybrid LSTM-GRU network). These architectures range from lightweight to complex models, which enable us to investigate how model complexity and structure impact network traffic patterns. We also evaluate multiple real-world datasets corresponding to different modalities (e.g., image classification datasets like CIFAR-10 and MNIST, and time-series datasets like UCI HAR for human activity recognition) to ensure that our proposed approach does not overfit to a single data domain. Furthermore, we test the fingerprinting attack under different federated aggregation algorithms: the standard Federated Averaging (FedAvg) \cite{mcmahan2017communication}, the weighted FedAvg \cite{pillutla2022robust}, and Federated Proximal (FedProx) \cite{yuan2022convergence}. By incorporating these heterogeneities into our analysis, we ensure that our proposed attack strategy generalizes across different FL systems. We summarize our key contributions as follows: 

\begin{itemize}[leftmargin=*]
    \item We uncover a novel side-channel fingerprinting attack in Federated Learning that infers the underlying deep learning model (e.g., CNN and RNN) by sniffing encrypted network traffic and without interfering with the FL process.

    \item We implement a realistic experimental testbed with heterogeneous devices and Wi-Fi connectivity [addresses \textbf{C1}]. We also consider attacks under varying sniffing times, to mimic a practical attack in FL deployments [addresses \textbf{C3}].
    
    \item We propose a late fusion strategy utilizing two meta-models, logistic regression (MetaLR) and gradient-boosted trees (MetaXGB), to synthesize flow-level and packet-level traffic characteristics and enhance the robustness of fingerprinting. To validate, we consider performance under both close-world and open-world settings [addresses \textbf{C2}]
    
    
\end{itemize}
\vspace{-0.05in}

\section{Related Work}\vspace{-0.05in}

\subsection{Security and Privacy in Federated Learning}\vspace{-0.05in}
Federated Learning (FL) was introduced to enhance data privacy by keeping user data local, but it is vulnerable to a range of privacy and integrity attacks. Adversaries can launch poisoning attacks—such as backdoor attacks that implant hidden triggers—by submitting manipulated model updates or conducting inference attacks to extract private information from gradient updates. Nasr et al. first proposed membership inference attacks in centralized deep learning, which were later extended to FL by exploiting shared gradients and model parameters~\cite{nasr2019comprehensive, zhu2025fedmia}. Inversion attacks, where adversaries reconstruct representative training samples from gradients, have also been demonstrated in FL scenarios~\cite{yang2025deep, chen2025deep}. Moreover,~\cite{mehnaz2022your,wang2022poisoning} showed that with black box access to federated models, attackers can find sensitive properties of the training set. While direct attacks—such as gradient leakage, model inversion, and poisoning—have been extensively studied in FL, side-channel vulnerabilities remain largely unexplored. In contrast to these active approaches, we propose a passive attack that infers model architecture using only network metadata, without accessing model parameters or gradients.

\subsection{Traffic Fingerprinting and Side-Channel Leakage}\vspace{-0.05in}
Traffic fingerprinting (TF) is a well-established technique in network security that analyzes patterns in network flows to infer hidden properties such as applications, protocols, or content—often without decrypting the traffic. In website fingerprinting, for instance, attackers can identify visited websites from encrypted HTTPS traffic using features like packet size and timing. Research showed that even Tor-protected traffic leaks enough metadata that enables inference via passive analysis~\cite{panchenko2016website},~\cite{oh2021gandalf}. Similar techniques were applied in device fingerprinting, where authors in \cite{laperdrix2020browser} demonstrated that browsers and devices can be identified through their network-level behavior. Other works targeted IoT environments, leveraging data transmissions and protocol-specific patterns to distinguish device types, usages, and models \cite{zhang2025physical,sheng2025network}. However, none of these methods explored whether deep learning architecture can be inferred from network metadata, particularly in an FL environment, and our work is the first to address this.
\vspace{-0.05in}

\subsection{Side-Channel Privacy in ML Systems}\vspace{-0.05in}
Side-channel attacks extract unintended information from machine learning (ML) systems by observing indirect signals such as timing, memory access, power consumption, or network activity. Nayak et al.~\cite{nayak2020data} provided a comprehensive overview of such vulnerabilities, categorizing how different stages of the ML pipeline can leak sensitive information through non-intrusive observations. In the context of federated learning (FL), side channels extend beyond gradient leakage and local data access to include system-level signals exploited without interfering with the FL protocol. For instance, Shokri et al.~\cite{shokri2017membership} showed that timing variations in training can reveal dataset size or indicate client-specific model complexity. Batina et al.~\cite{batina2019csi} demonstrated that power consumption traces from edge devices can reflect computational workload. Recent research~\cite{zhang2023deep,chabanne2021side} explored whether model architecture can be extracted from side-channel information like power usage and execution timing in centralized ML systems. While prior side-channel attacks primarily focused on centralized ML systems—exploiting hardware-level signals such as power usage or timing through probing mechanisms—our work targets federated learning and demonstrates that model architecture can be inferred purely from network-layer metadata.



\vspace{-0.05in}

\section{Proposed Threat Model and System Overview}\vspace{-0.05in}
This section outlines the threat model, the system model, and the experimental design.\vspace{-0.05in}

\begin{figure}[t]
    \centering
    \includegraphics[width=\columnwidth]{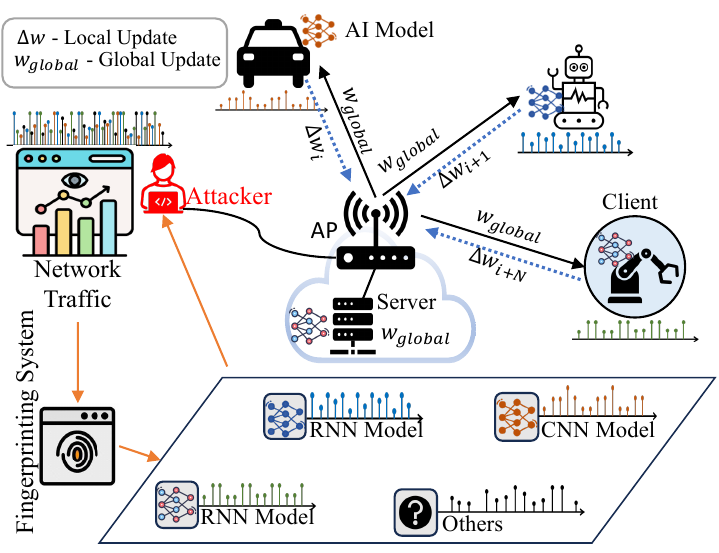}
    \caption{The proposed threat model.}
    \label{fig:at_model}
    \vspace{-0.3in}
\end{figure}

\subsection{Threat Model}\vspace{-0.05in}
In this research, we consider a passive adversary in FL settings who aims to infer the deep learning model architecture being trained on different clients' devices by eavesdropping on the side channel network information. The adversary is not a participant in the FL training process and does not contribute model updates or take part in aggregation. They do not interfere with or alter the FL process in any way (such as gradient poisoning or malicious behavior); instead, they remain entirely passive. Fig.\ref{fig:at_model} illustrates the threat model of this system. We consider a realistic scenario where the attacker passively monitors wireless traffic by residing at the access point (AP), which is assumed to be wired to the federated server. This setup reflects common edge FL deployments, where client devices connect to the server via an observable wireless gateway\cite{wang2019edge}. Such an adversary may be a compromised router, a passive eavesdropper on the wireless link, or even an honest-but-curious edge node with network-level visibility, but is not a participant in the FL training process. As the AP can observe all traffic between clients and the server, the attacker can isolate communication flows and identify individual clients through side-channel metadata.

\textbf{Adversary Capabilities:} We assume the adversary has the following capabilities, considering it can only access the side channel information through eavesdropping.
\begin{itemize}
    \item Traffic Sniffing: Since all upstream and downstream communication passes through the AP, the attacker can capture the wireless traffic data during training. Therefore, clients associated with specific network traces can also be identifiable using corresponding MAC addresses. 
    \item Metadata Extraction: Although packet contents are encrypted, an adversary can access side-channel metadata (e.g., packet sizes, transmission direction (uplink/downlink), timestamps, inter-arrival intervals, and burst patterns), which are sufficient to reconstruct the traffic pattern without accessing payloads. 
\end{itemize}

\textbf{Adversary Limitation:} The adversary operates passively and does not inject, modify, or drop packets. It does not perform active probing or interfere with the training protocol in any manner. In addition, it cannot decrypt packet payloads or access internal FL information, e.g., model parameters, gradients, or training data. Their visibility is strictly limited to side-channel metadata at the wireless access point.

We assume that each client trains only one FL model per session, aligning with standard deployment practices on mobile and edge devices, where resource constraints typically limit concurrent model training. This helps simplify flow separation by reducing potential traffic overlap. We also assume that MAC address randomization is disabled, allowing the attacker to associate a single MAC address with each client.

Moreover, the attacker does not have access to higher-layer protocol information (e.g., TCP sequence numbers). We assume that each client trains only one FL model per session, which aligns with common deployment practices on mobile and edge devices where resource constraints typically limit concurrent model training. This simplifies flow separation by reducing potential traffic overlap. We also assume that MAC address randomization is disabled, allowing the attacker to associate a single MAC address with each client. 

\textbf{Attack Goal:} The attacker aims to fingerprint the model architecture used by each client during FL training. Specifically, the attacker deanonymizes CNN and RNN architectures; traffic associated with other model types is labeled as unknown. For each client device, the adversary aims to classify the DL model being trained (e.g., ResNet18, BiLSTM) by analyzing the extracted traffic metadata. This task is formulated as a multi-class classification problem, where the input features are side-channel observations (e.g., packet size, inter-arrival time) and the output is the predicted model architecture (e.g., CNN, RNN). A successful attack allows the adversary to associate each communication pattern with its underlying model type. \vspace{-0.05in}

\begin{figure*}[t]
    \centering
    \includegraphics[width=\textwidth]{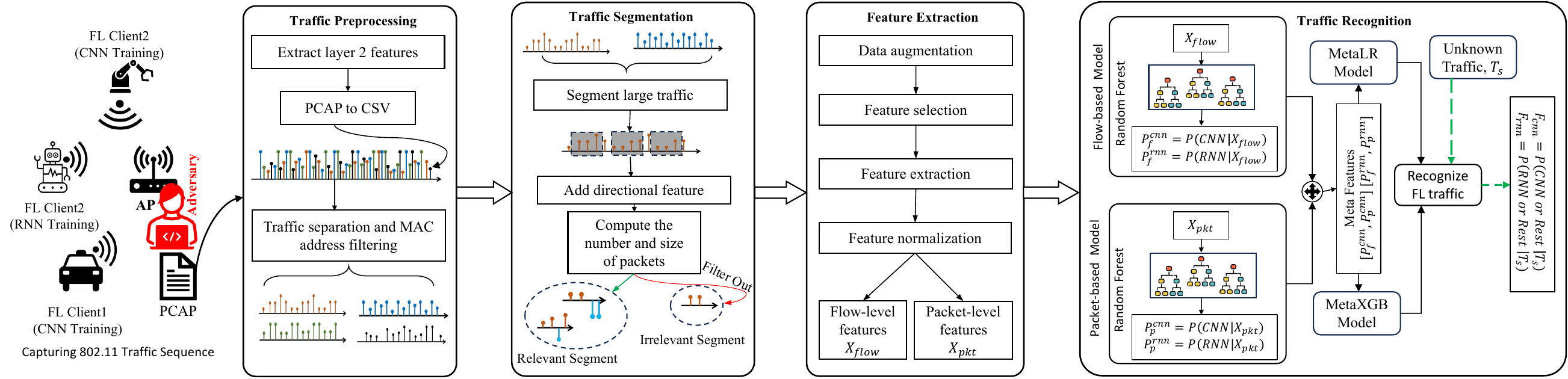}\vspace{-0.05in}
    \caption{System workflow of FLARE.}
    \label{fig:sys_model}
    \vspace{-0.3in}
\end{figure*}

\subsection{System Model}\vspace{-0.05in}
To evaluate the feasibility of the proposed fingerprinting strategy in FL, we develop a multi-stage testbed that captures realistic network-layer characteristics in diverse training conditions. This section outlines our FL framework, communication process, experimental setup, and data collection methodology. 

\subsubsection{Federated Learning Framework}
We implement a synchronous FL framework involving a server and $K$ clients communicating via Wi-Fi. In each round $n$, the server broadcasts the current global model $w^n \in \mathbb{R}^d$ to all selected clients. Each client $k$ performs local training on private data $\mathcal{D}_k$ and returns a model update $\Delta w_k^n$. The server aggregates updates via an aggregation strategy $\mathcal{A}$ to produce the next global model:\vspace{-0.1in}

\begin{equation}
w^{n+1} = \mathcal{A}\left( \{ \Delta w_k^n \}_{k \in K_n} \right).
\end{equation}\vspace{-0.2in}

We consider three aggregation strategies: FedAvg, where the server takes a weighted average of client updates; weighted FedAvg, where the average is calculated by considering the data proportion to emphasize scenarios with unbalanced data distribution across clients; and FedProx, where clients use proximal regularization to limit global deviations. In each training round, clients run a few local epochs, and then the updates are aggregated on the server \textbf{[C2]}.

Moreover, here, the model update size scales with the architecture. Let $\theta$ denote the number of trainable parameters. Assuming 32-bit floats, the total payload per update is approximately $S_k^n = 4 \cdot \theta \quad \text{(in bytes)}$. This variation of payload size differs in different model families (e.g., CNN vs RNN), which is one of our key hypotheses for fingerprinting traffic.

\subsubsection{Models (Architecture Scope)}

In this research, our goal is to distinguish \textbf{broad architecture families}. In order to do that, our architectures fall into two broad categories: CNN-based and RNN-based. The CNN-based models include the custom CNN-based models and pre-trained models like ResNet18, MobileNetV2, and DenseNet. On the other hand, RNN-based models include the Vanilla RNN model, the LSTM-based model, the BiLSTM-based model, the GRU-based models, and the hybrid LSTM-GRU model \textbf{[C2]}. All RNN models are trained on sequence data. As there are no pre-trained RNN-based models available, we therefore choose to use custom-made RNN models and several structures from various research studies on time series analysis.

\subsubsection{FL Dataset}
Several DL models on different DL families trained on domain-appropriate datasets, and these datasets are widely used in federated learning research \textbf{[C2]}, including:
\begin{itemize}
    \item \textit{Image datasets:}  \textbf{CIFAR-10} (60k 32×32 color images in 10 classes),  \textbf{Fashion-MNIST} (70k 28×28 grayscale images of clothing categories),  \textbf{MNIST} (handwritten digits),  \textbf{SVHN} (Street View House Numbers—another image dataset of digits), and  \textbf{STL-10}.
    \item \textit{Sequence datasets:}  \textbf{UCI HAR} (the UCI Human Activity Recognition dataset from smartphone sensors), which is a multivariate time-series classification task; a  \textbf{Sunspots} time series dataset (regression problem of sunspot activity over time); and ECG Dataset.
\end{itemize}
\vspace{-0.05in}
Each model-dataset pair is trained using the same FL pipeline. We also ensure the local dataset is distributed in a non-IID manner, which represents realistic FL scenarios. The key reason for using multiple datasets along with different models is to incorporate diversity, which ensures that the fingerprinting classifier generalizes across datasets and is not biased.
\vspace{-0.05in}

\subsection{Experimental Design}

\subsubsection{Deployment Setup}
In our testbed, we deployed our FL system using heterogeneous edge devices connected over private Wi-Fi. In our setup, we use a Core i7 desktop with an RTX 3060 GPU as a federated server. This server is wiredly connected with a Wi-Fi access point (AP). Then we chose three different clients for our system, such as (1) Jetson Orin (embedded with Nvidia ARM GPU), a laptop with a GTX 990 GPU, and a MacBook Pro with an M1 chip. The server and Orin run a Linux system, whereas the laptop runs on Windows OS, and the MacBook runs on macOS. This heterogeneity means that the computation speed and network interface of each client differ. We ensured all clients started each round simultaneously. To achieve this, we implement synchronization between the server and clients to allow the server to wait for all clients to finish a round before receiving updates from other clients. Moreover, different hardware can introduce variations in timing (e.g., the Windows laptop takes longer per epoch than the Jetson), which can affect when their updates are sent. We considered this design to overcome \textbf{C1}.
\begin{figure}[ht!]
    \centering
    \includegraphics[width=1\columnwidth]{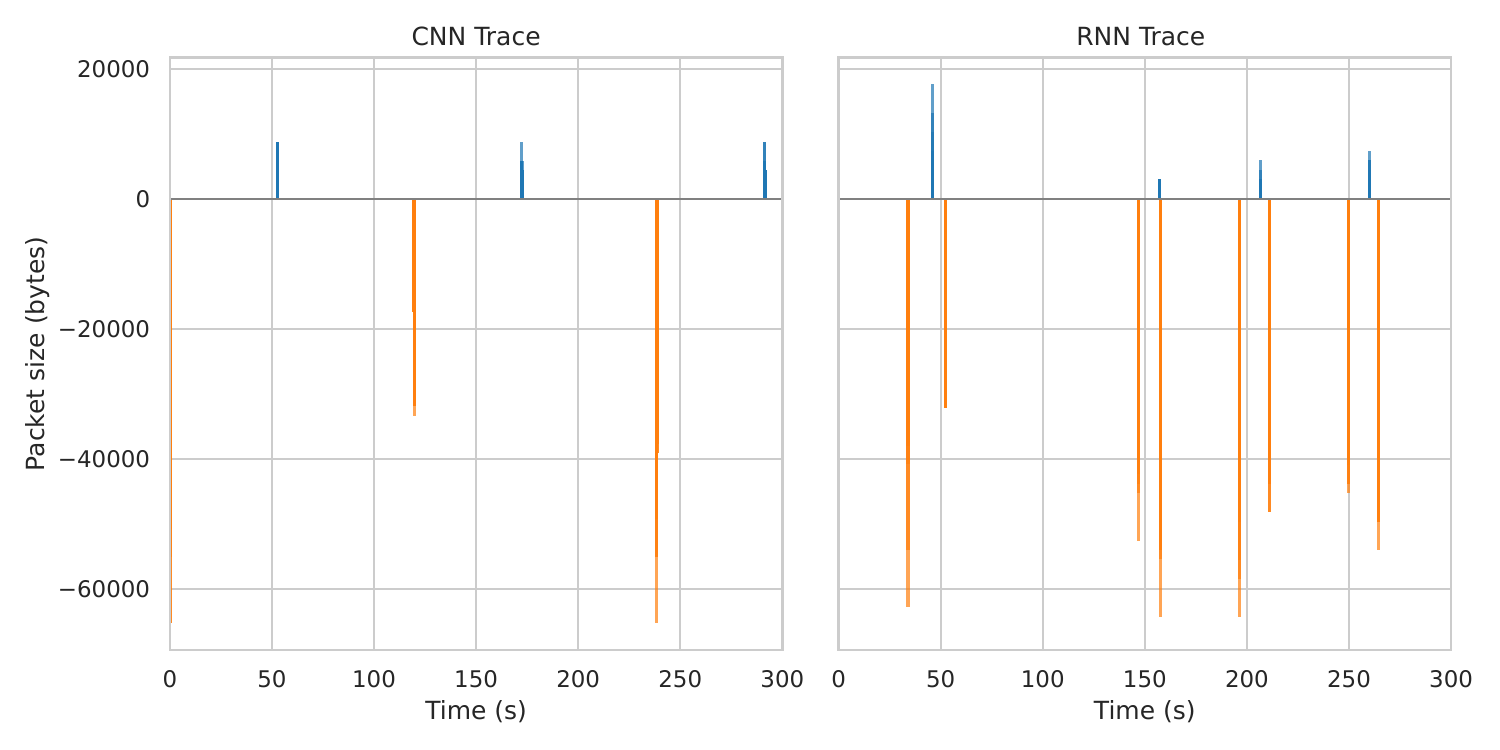}\vspace{-0.15in}
    \caption{Traffic of CNN vs RNN training (Blue: Uplink, Orange: Downlink).}
    \label{fig:trace}\vspace{-0.25in}
\end{figure}
\vspace{-0.05in}

\vspace{-0.05in}
\section{FLARE: Proposed Fingerprinting Framework}\vspace{-0.05in}

We refer to our proposed fingerprinting framework as FLARE (Federated Learning Architecture REconnaissance), which passively infers the deep learning model architecture of FL clients by analyzing wireless traffic patterns. \vspace{-0.05in}

\subsection{FLARE Workflow}
As shown in Fig.~\ref{fig:sys_model}, the proposed fingerprinting framework operates in three stages: (1) Traffic Data Collection and Preprocessing, (2) Traffic Segmentation, (3) Feature Extraction, and  (4) Traffic Recognition.

\subsubsection{Traffic Data Collection and Processing}
In our FL testbed, all clients communicate with the server using the IEEE 802.11ac protocol, while the server is connected via Ethernet. We assume a passive adversary with privileged visibility at the access point (AP), which we emulate by capturing wireless traffic at the AP using Wireshark in monitor mode. The raw traffic is recorded in PCAP format for each training session, covering the entire duration until the model reaches satisfactory accuracy. Hence, the number of packets varies depending on model size, training speed, and client hardware.

We extract side-channel metadata from the captured traces, including packet size, transmission direction (uplink/downlink), and inter-arrival time (IAT). The traffic reflects significant heterogeneity due to variations in model architectures, datasets, and device types, mirroring real-world FL deployments. For analysis, PCAP traces are converted to structured CSV format using Python. Since the adversary knows the AP's MAC address, MAC filtering isolates traffic per client. Although these clients may or may not participate in FL training, filtering enables accurate per-device flow separation.

\subsubsection{Traffic Segmentation}
Assuming the adversary can start eavesdropping at any point in the FL training phase, we segment the traffic into fixed-duration windows of 250–300 seconds. This duration reflects a realistic observation window—sufficient to capture meaningful communication patterns while remaining practical for passive monitoring \textbf{[C3]}.
Fig.~\ref{fig:trace} shows example traffic segments for CNN and RNN training. CNN traffic exhibits sparse uplink and large downlink packets, while RNN traces display more frequent and periodic bidirectional communication. These differences provide visual evidence of architectural differences based on traffic dynamics.

\subsubsection{Feature Extraction}
We hypothesize that different deep learning architectures (e.g., CNNs vs. RNNs) produce statistically distinct traffic patterns during training. To effectively capture these differences, we extract two complementary categories of features from each traffic window: flow-level features and packet-level features. These perspectives provide coarse-grained and fine-grained views of communication behavior, enabling more robust and discriminative model fingerprinting.

\textbf{Flow-Level Features ($\mathbf{X}_\text{flow}$):}
Flow-level features capture long-term, coarse-grained patterns of a traffic segment over a fixed window. These features capture global temporal dynamics and statistical summaries of the traffic. Specifically, we extract Packet rate statistics (e.g., mean, max, min, median, and standard deviation of packet counts per second); Directional packet size statistics (e.g., mean, max, min, variance, standard deviation, median absolute deviation, skewness, kurtosis, and 10th to 90th percentile). These features are important because they reflect the training dynamics of the underlying model. As shown in Fig.~\ref{fig:trace}, CNN traffic appears as sparse bursts due to large updates, while RNN traffic is more periodic—patterns effectively captured by flow-level features.

\textbf{Packet-Level Features ($\mathbf{X}_\text{pkt}$):}
Packet-level features are designed to capture localized structural patterns within each traffic window. These features differ from flow-level statistics in that they preserve localized variations rather than summarizing aggregated trends. First, we extract a packet length histogram, which represents the distribution of packet sizes across fixed-size bins. This captures the overall shape and variation in packet lengths, revealing architecture-specific behaviors such as burst sizes or dominant packet types. Unlike directional packet size statistics used in flow-level features—which summarize aggregated trends (e.g., the mean or variance of uplink and downlink packet sizes)—the histogram reflects the frequency with which packets of different sizes occur, independent of direction. In addition, we include edge statistics by recording the size and IAT of the first and last packets within each window. Although the attacker may join at any point during training, the first and last packets within each observed window still reflect architecture-specific traffic patterns—such as transient bursts in CNNs or periodic, low-volume activity in RNNs. These edge features capture the local communication structure that complements the global features.

\subsubsection{Traffic Recognition}
This stage involves three components: (1) independent modeling of flow-level and packet-level traffic features, (2) fusion of predictions into meta-features, and (3) final classification via a fusion meta-model. We formulate the classification task as a binary decision problem. However, directly modeling CNN vs. RNN traffic limits the system's applicability to closed-world settings and cannot effectively distinguish non-FL or unrelated traffic. To address this, we adopt a one-vs-rest approach, where the system learns to detect whether a traffic segment belongs to a specific architecture family (CNN or RNN). This formulation improves generalization in open-world scenarios, where new model types may appear. It also enhances scalability, as new model families (e.g., Transformers) can be incorporated as additional one-vs-rest classifiers without retraining the entire system. Accordingly, we implement two parallel pipelines: one for CNN vs. Rest and another for RNN vs. Rest.

\textbf{Independent Modeling:}
Independently train two base classifiers based on the Random Forest classifier. This ensemble method contains multiple decision trees trained using bagging and random feature selection. It can handle non-linear data and give intrinsic feature importance, making it suitable for fingerprinting tasks \cite{breiman2001random}. Each model independently predicts the likelihood of the traffic originating from a CNN architecture vs. Rest and an RNN architecture vs Rest.
\begin{itemize}[leftmargin=*]
    \item \textbf{Flow-based model:} Random Forest model trained on flow-level features $\mathbf{X}_\text{flow}$, yielding a meta prediction $P_f^{cnn} = P(\text{CNN} \mid \mathbf{X}_{flow})$ and $P_f^{rnn} = P(\text{RNN} \mid \mathbf{X}_{flow})$ for CNN and RNN classes, respectively.
    
    \item \textbf{Packet-based model:} Random Forest model trained  on packet-level features  $\mathbf{X}_\text{pkt}$, yielding a meta prediction score $P_p^{cnn} = P(\text{CNN} \mid \mathbf{X}_{pkt})$ and $P_p^{rnn} = P(\text{RNN} \mid \mathbf{X}_{pkt})$ for CNN and RNN classes, respectively.
\end{itemize}

\textbf{Meta-Fusion Classification:}
In our second stage, we adopt a \textit{late fusion} strategy, where the decisions of independently trained base models are aggregated at a higher level. This design choice is motivated by the heterogeneous nature of flow-level and packet-level characteristics, which often exhibit distinct statistical properties. Training separate models for each view allows them to specialize and learn modality-specific patterns more effectively \cite{baltruvsaitis2018multimodal,ngiam2011multimodal}. Moreover, late fusion enhances robustness in scenarios where one modality may be noisy or degraded, as each model contributes independently.

To integrate the predictive knowledge from both the flow and packet models, we construct a meta-feature vector:\vspace{-0.05in}

\begin{equation}
\mathbf{x}^{(\ell)}_{\text{meta}} = \left[ P_f^{(\ell)},\; P_p^{(\ell)} \right], \quad \ell \in \{\text{cnn},\; \text{rnn}\}.
\tag{2}
\end{equation}

where $P_f$ and $P_p$ represent the predicted probabilities from the flow and packet classifiers, respectively. This vector is fed into a fusion classifier, whose role is to synthesize the modality-specific insights and produce a unified prediction.

We evaluate two different meta-classifiers for the fusion stage: logistic regression (MetaLR) and gradient boosted trees (MetaXGB). Since this is the first work to explore model architecture fingerprinting in FL using side-channel traffic, there is no prior benchmark or standard fusion model to compare against. Our goal, therefore, is to examine how two fundamentally different classification paradigms behave in this context: MetaLR offers a simple, interpretable linear decision boundary, while MetaXGB represents a powerful non-linear ensemble method capable of modeling complex interactions. We denote the fusion meta-classifier as $g_\theta$, where $\theta$ represents the learned parameters by MetaLR or MetaXGB. Given meta-features constructed from flow-level and packet-level predictions, $g_\theta$ outputs the final architecture prediction. By evaluating both, we aim to demonstrate the robustness and adaptability of our multi-perspective fusion strategy across different classifiers. Given an unkown traffic segment $t$, the final fingerprint prediction is computed as:\vspace{-0.1in}

\begin{equation}
F_\ell = g_\theta^{(\ell)}(\mathbf{x}^{(\ell)}_{\text{meta}}) \in \{(\text{CNN}, \text{Rest}), (\text{RNN}, \text{Rest})\}.
\tag{3}
\end{equation}
\noindent
where \( \ell \in \{\text{CNN}, \text{RNN}\} \) denotes the target architecture class.
\noindent
Here, \( F_\ell \) denotes the meta-classifier associated with class \( \ell \). Each classifier maps the meta-level feature vector \( \mathbf{x}_{\text{meta}} \) to a binary prediction between the target architecture (CNN or RNN) and all other model types.

This layered approach not only improves classification performance but also enhances interpretability and modularity, as each base model can be inspected and tuned independently~\cite{ramachandram2017deep}. Furthermore, for training the fingerprinting model, all models are trained in a supervised manner, where all traffic traces are collected and labeled. We utilize stratified cross-validation and grid search for hyperparameter tuning. To formalize the objective of our fingerprinting framework, we express the learning task as an optimization problem that integrates the outputs of the flow-based and packet-based classifiers into a unified prediction through a fusion meta-model. This formulation models how the flow-level and packet-level predictions are combined in the final meta-classifier, which is designed to leverage the complementary information extracted during the earlier stages. Prior studies (e.g., \cite{ramachandram2017deep}, \cite{baltruvsaitis2018multimodal}) highlight late fusion's robustness in multimodal and side-channel learning contexts, especially when feature domains are heterogeneous or differently scaled.

\subsection{FLARE Training Algorithm}
We summarize the end-to-end training process of FLARE in Algorithm~\ref{algo:flare_training}. Given a dataset $\mathcal{D} = {(T_i, y_i)}_{i=1}^M$ containing $M$ traffic traces and their corresponding model architecture labels, we discard any trace lacking packets above a threshold size $\tau < 66$ bytes. The threshold is chosen to exclude control or idle traffic and retain training-relevant activity. For the remaining traces, we extract two views of the traffic—$\mathbf{x}_i^{(f)}$ representing flow-level features and $\mathbf{x}_i^{(p)}$ for packet-level features. These features capture both temporal and structural patterns in the FL traffic. Once features are extracted, two independent classifiers $h_f$ and $h_p$ are trained to process the flow-level and packet-level information, respectively. Then, their outputs are concatenated to form a meta-feature vector $\mathbf{z}_i = [h_f(\mathbf{x}_i^{(f)}), h_p(\mathbf{x}_i^{(p)})]$, which is passed to a fusion model $g_\theta$. The final model prediction is computed as $F_i = g_\theta(\mathbf{z}_i)$.

In order to train our fingerprinting pipeline in an offline manner, we formulate our problem as an optimization problem that minimizes the following fusion loss: 

\begin{small}
\begin{equation}
\begin{aligned}
\textbf{minimize} \quad & \mathcal{L}_{\text{fusion}} = \frac{1}{M} \sum_{i=1}^{M} 
\mathcal{L} \left( g_\theta^{(\ell)}([ h_f(x_i^f), h_p(x_i^p) ]),\; y_i \right), \\
\textbf{subject to} \quad & h_f: \mathbb{R}^{x_i^f} \rightarrow [0, 1], \quad h_p: \mathbb{R}^{x_i^p} \rightarrow [0, 1], \\
& g_\theta^{(\ell)}: \mathbb{Z}^2 \rightarrow \{0,1\}, \quad \theta \in [CNN, RNN].
\end{aligned}
\label{eq:fusion_optimization}
\end{equation}
\end{small}

Here, $\mathcal{L}(\cdot, \cdot)$ denotes the binary cross-entropy loss and $g_\theta^{(\ell)}$ is either logistic regression (MetaLR) or XGBoost (MetaXGB), depending on the selected fusion strategy. 

\begin{algorithm}[htb]
\caption{FLARE: End-to-End Training Mechanism}
\label{algo:flare_training}
\begin{algorithmic}
\STATE \textbf{Input:} Labeled traces $\mathcal{D} = \{(T_1, y_1), \dots, (T_M, y_M)\}$
\STATE \textbf{Output:} Trained models $h_f$, $h_p$, and $\{g_\theta^{(\ell)}\}_{\ell \in \{\text{cnn}, \text{rnn}\}}$
\vspace{0.5em}
\STATE \textit{// Step 1: Feature Extraction}
\FOR{each $T_i \in \mathcal{D} : \max(\texttt{PacketSize}(T_i)) > \tau$}
    \STATE $\mathbf{x}_i^{(f)}, \mathbf{x}_i^{(p)} \gets \texttt{FeatureExtractor}(T_i)$
\ENDFOR
\vspace{0.5em}
\STATE \textit{// Step 2: Architecture-Specific Fusion Model Training}
\FOR{each $\ell \in \{\text{cnn}, \text{rnn}\}$}
    \FOR{$i = 1$ to $M$}
        \STATE $P_f^{(\ell)} \gets h_f(\mathbf{x}_i^f), \quad P_p^{(\ell)} \gets h_p(\mathbf{x}_i^p)$
        \STATE $\mathbf{x}_{meta}^{(\ell)} \gets [P_f^{(\ell)}, P_p^{(\ell)}]$
        \STATE $F_i^{(\ell)} \gets g_\theta^{(\ell)}(\mathbf{x}_{meta}^{(\ell)}), \text{where } \theta \in \{\text{MetaXGB}, \text{MetaLR}\}$
    \ENDFOR
    \STATE $g_\theta^{(\ell)} \gets \arg\min_{\theta \in \Theta} \frac{1}{M} \sum_{i=1}^{M} \mathcal{L}(F_i^{(\ell)}, y_i)$
\ENDFOR
\RETURN $h_f$, $h_p$, $\{g_\theta^{(\ell)}\}_{\ell \in \{\text{cnn}, \text{rnn}\}}$
\end{algorithmic}
\end{algorithm}\vspace{-0.1in}

\vspace{-0.05in}

\section{Evaluation and Result Analysis}
\label{sec:evaluation}
\begin{table*}[ht]
\centering
\caption{Performance Analysis of FLARE in open-world and Closed-world settings (mean ± standard deviation)}
\label{tab:my-table}
\resizebox{\textwidth}{!}{
\begin{tabular}{cccccccccc}
\hline
\multirow{2}{*}{Scenario} & \multirow{2}{*}{Class} & \multirow{2}{*}{Metric} & \multicolumn{3}{c}{MetaLR} &  & \multicolumn{3}{c}{MetaXGB} \\ \cline{4-6} \cline{8-10} 
 &  &  & Flow & Packet & FLARE (Fusion) &  & Flow & Pakcet & FLARE (Fusion)\\ \hline
\multirow{6}{*}{Close World} & \multirow{3}{*}{CNN vs Rest} & Precision & 0.989 ± 0.018 & 0.969 ± 0.042 & \textbf{0.985} ± 0.017 &  & 0.995 ± 0.011 & 0.965 ± 0.033 & \textbf{0.985} ± 0.020 \\
 &  & Recall & 0.938 ± 0.025 & 0.835 ± 0.065 & \textbf{1.000} ± 0.000 &  & 0.949 ± 0.023 & 0.804 ± 0.035 & \textbf{1.000} ± 0.000 \\
 &  & F1-Score & 0.963 ± 0.016 & 0.894 ± 0.024 & \textbf{0.992} ± 0.008 &  & 0.971 ± 0.010 & 0.876 ± 0.016 & \textbf{0.992} ± 0.010 \\ \cline{2-10} 
 & \multirow{3}{*}{RNN vs Rest} & Precision & 0.976 ± 0.024 & 0.948 ± 0.033 & \textbf{1.000} ± 0.000 &  & 0.987 ± 0.027 & 0.951 ± 0.060 & \textbf{1.000} ± 0.000 \\
 &  & Recall & 0.925 ± 0.103 & 0.660 ± 0.161 & 0.962 ± 0.042 &  & 0.912 ± 0.094 & 0.566 ± 0.180 & \textbf{0.963} ± 0.050 \\
 &  & F1-Score & 0.945 ± 0.051 & 0.766 ± 0.111 & \textbf{0.980} ± 0.022 &  & 0.946 ± 0.057 & 0.688 ± 0.148 & \textbf{0.980} ± 0.027 \\ \hline
\multirow{6}{*}{Open World} & \multirow{3}{*}{CNN vs Rest} & Precision & 1.000 ± 0.000 & 0.867 ± 0.267 & \textbf{0.850} ± 0.122 &  & 1.000 ± 0.000 & 0.567 ± 0.361 & 0.838 ± 0.096 \\
 &  & Recall & 0.733 ± 0.327 & 0.467 ± 0.163 & \textbf{1.000} ± 0.000 &  & 0.875 ± 0.125 & 0.479 ± 0.291 & \textbf{1.000} ± 0.000 \\
 &  & F1-Score & 0.800 ± 0.245 & 0.587 ± 0.185 & \textbf{0.914} ± 0.070 &  & 0.929 ± 0.071 & 0.510 ± 0.305 & 0.909 ± 0.054 \\ \cline{2-10} 
 & \multirow{3}{*}{RNN vs Rest} & Precision & 1.000 ± 0.000 & 0.783 ± 0.194 & \textbf{1.000} ± 0.000 &  & 1.000 ± 0.000 & 0.688 ± 0.410 & \textbf{1.000} ± 0.000 \\
 &  & Recall & 0.833 ± 0.211 & 0.700 ± 0.267 & \textbf{0.800} ± 0.163 &  & 0.875 ± 0.217 & 0.479 ± 0.291 & 0.792 ± 0.125 \\
 &  & F1-Score & 0.893 ± 0.137 & 0.718 ± 0.203 & \textbf{0.880} ± 0.098 &  & 0.917 ± 0.144 & 0.554 ± 0.323 & 0.879 ± 0.074 \\ \hline
 \end{tabular}
 }\vspace{-0.2in}
\end{table*}

We evaluate the effectiveness of the proposed FLARE framework under both open-world and closed-world scenarios. In each case, we train binary \textit{One-vs-rest} classifiers (CNN and RNN) using flow-level and packet-level features independently, and in combination through a late fusion model.
\vspace{-0.05in}

\subsection{Evaluation Metrics}
We use three standard classification metrics. \textbf{Precision}: The proportion of samples predicted as class $C$ that are truly from $C$. \textbf{Recall}: The proportion of actual class $C$ samples correctly identified. \textbf{F1-Score}: Harmonic mean of precision and recall, providing a balanced performance summary.
\vspace{-0.05in}
\subsection{Evaluation Setup}
Each meta-classifier is trained using 5-fold stratified cross-validation. During testing, we evaluate performance on a separate test set derived from multiple deployment scenarios. Results are reported using standard classification metrics—precision, recall, and F1-score—presented as \textbf{(mean ± standard deviation)} across test samples. To assess the consistency and statistical significance of our results, we also compute 95\% confidence intervals across five independent runs for each model configuration using the Student’s t-distribution (t = 2.776, n = 5, df = 4). The observed intervals between fusion and baseline models suggest that performance improvements are statistically meaningful.



\subsection{FLARE’s Fingerprint Performance}\vspace{-0.05in}
We evaluate the fingerprinting accuracy of FLARE under both closed-world and open-world settings using two binary classification tasks: CNN-vs-Rest and RNN-vs-Rest. For each scenario, we compare the performance across three feature perspectives: flow-only, packet-only, and fusion (FLARE), and two meta-classifiers: Logistic Regression (MetaLR) and XGBoost (MetaXGB). Table~\ref{tab:my-table} summarizes the precision, recall, and F1-score (mean ± std) across all configurations.

\subsubsection{Closed-World Scenario}
FLARE achieves highly reliable fingerprinting performance in the closed-world setting, where all model classes are known during training and testing. Fusion-based classifiers consistently outperform individual flow-only or packet-only models in both tasks. For CNN-vs-Rest, MetaLR fusion achieves a precision of 0.985 ± 0.017, recall of 1.00 ± 0.00, and F1-score of 0.992 ± 0.008. Flow-only and packet-only F1-scores are 0.963 ± 0.016 and 0.894 ± 0.024, respectively, confirming the benefit of combining statistical and fine-grained traffic features.

The RNN-vs-Rest task shows similar trends. MetaXGB fusion yields 1.000 ± 0.000 precision, 0.963 ± 0.050 recall, and 0.980 ± 0.027 F1-score. In contrast, the packet-only classifier achieves a lower F1-score of 0.688 ± 0.148, reflecting its sensitivity to short-duration fluctuations and burst-level variability. Flow-only models perform better (0.946 ± 0.057), but still fall short of fusion in accuracy and consistency.

\textbf{Insight:} In closed-world settings, fusion yields high accuracy and stability, particularly for CNN. Although flow-level features alone are effective, fusion significantly improves robustness, especially for RNNs, overcoming \textbf{C2, C3}.

\subsubsection{Open-World Scenario}
Open-world evaluation introduces previously unseen architectures during testing, simulating real-world deployments. Despite this increased difficulty, fusion classifiers maintain strong performance. For CNN-vs-Rest, MetaLR fusion achieves an F1-score of 0.914 ± 0.070, outperforming packet-only (0.587 ± 0.185) and closely matching flow-only (0.929 ± 0.071). The perfect recall (1.000 ± 0.000) indicates successful detection of all CNN traffic, while the slightly reduced precision (0.850 ± 0.122) reflects false positives from unseen models mimicking CNN-like behavior.

For RNN-vs-Rest, MetaXGB fusion attains 1.00 ± 0.00 precision, 0.792 ± 0.125 recall, and 0.879 ± 0.074 F1-score. The corresponding packet-only F1-score is notably lower (0.554 ± 0.323), while flow-only achieves 0.917 ± 0.144. Once again, fusion demonstrates improved stability with lower variance, suggesting stronger generalization under model uncertainty.

\textbf{Insight:}  In open-world settings, fusion offers a consistent performance advantage by reducing variance and suppressing packet-level instability. Flow-only models remain relatively robust, but fusion is more resilient to previously unseen traffic patterns that confuse packet-based classifiers. Notably, FLARE maintains high sensitivity to true positives, achieving perfect recall in several cases. This demonstrates its ability to detect target architectures even under unfamiliar conditions. These results affirm FLARE’s robustness in realistic deployment scenarios, hence overcoming \textbf{C2, C3}.
\vspace{-0.05in}
\begin{figure}[htb!]
    \centering
    \includegraphics[width=\linewidth]{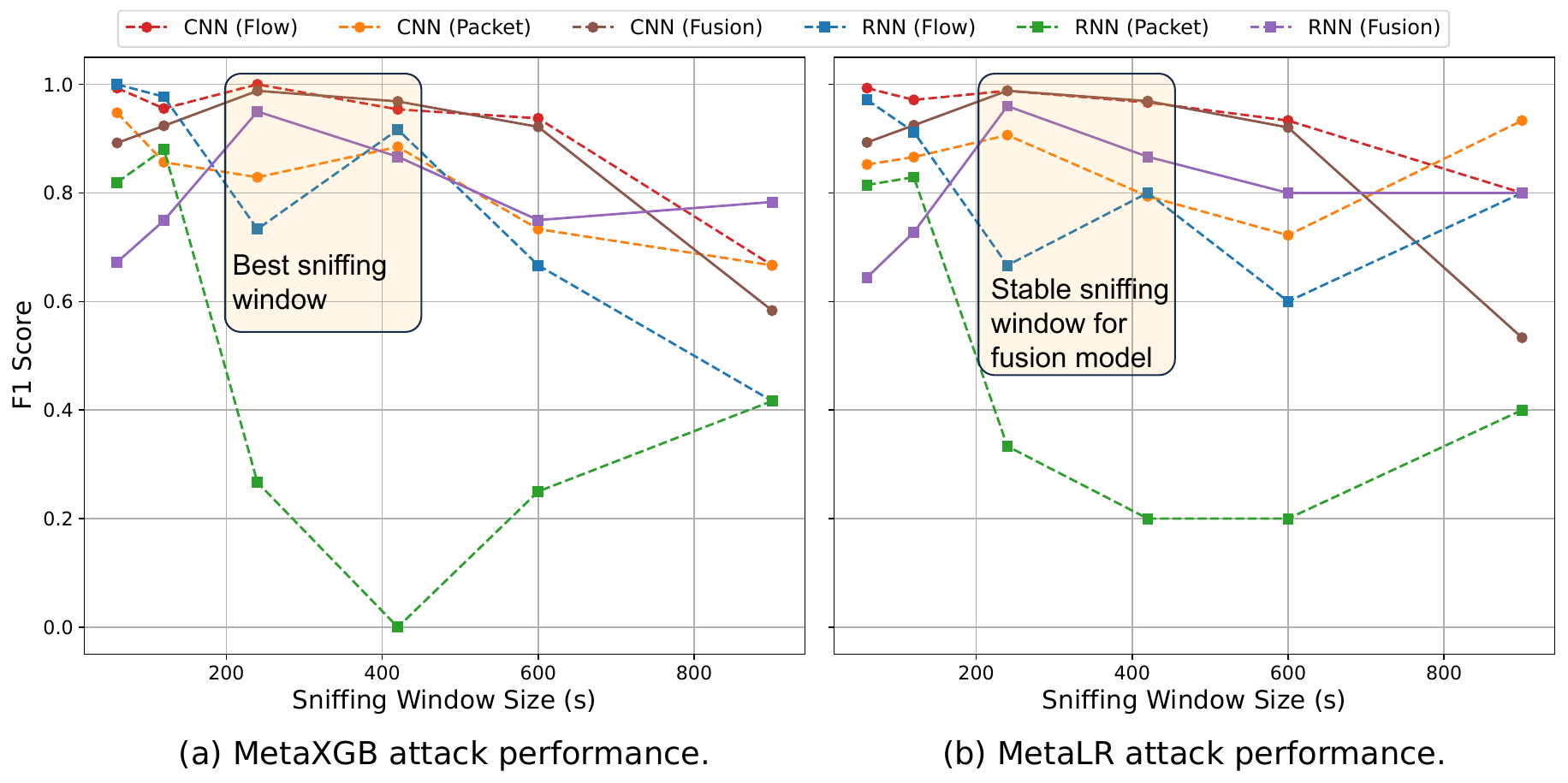}\vspace{-0.05in}
    \caption{Attack time vs. performance. 
    (a) Meta-XGB and
    (b) Meta-LR performance during varying attack windows.}
    \label{fig:attack_time}\vspace{-0.2in}
\end{figure}
\subsubsection{Impact of Observation Window Length}
We evaluate how the sniffing window affects model fingerprinting by comparing F1 scores across different sniffing durations for both MetaLR and MetaXGB classifiers (Fig.~\ref{fig:attack_time}). As the window increases from 60s to 240s, F1-scores generally improve due to richer statistical context, particularly for sparse or periodic RNN traffic. Between 240s to 420s, the performance remains relatively stable. Based on this trend, we identify 250–420s as a practical window range for effective attacks. After 420s, performance begins to degrade, as longer windows include idle intervals and multiple training rounds. However, until the 600s, the attack performance is close to 80\%. Notably, for RNN-vs-Rest classification, the fusion model at 600s window even outperforms other model perspectives. This causes feature smoothing, where distinctive patterns are averaged out, and the degradation is most evident in packet-level models. Beyond 600s, the degradation becomes more severe. 

\textbf{Insight:} Fingerprinting accuracy is highly sensitive to the chosen observation window. A window size of 250–600s offers a robust operational range for attackers, while fusion-based models provide consistent performance even under variable traffic durations. Hence, this result supports FLARE’s practicality and stability in real-world scenarios with uncertain observation opportunities, overcoming \textbf{C3}.\vspace{-0.05in}

\subsection{Feature Separability Analysis}
\begin{figure}[h]
\centering
\subfigure[]
{\includegraphics[width=0.24\textwidth]{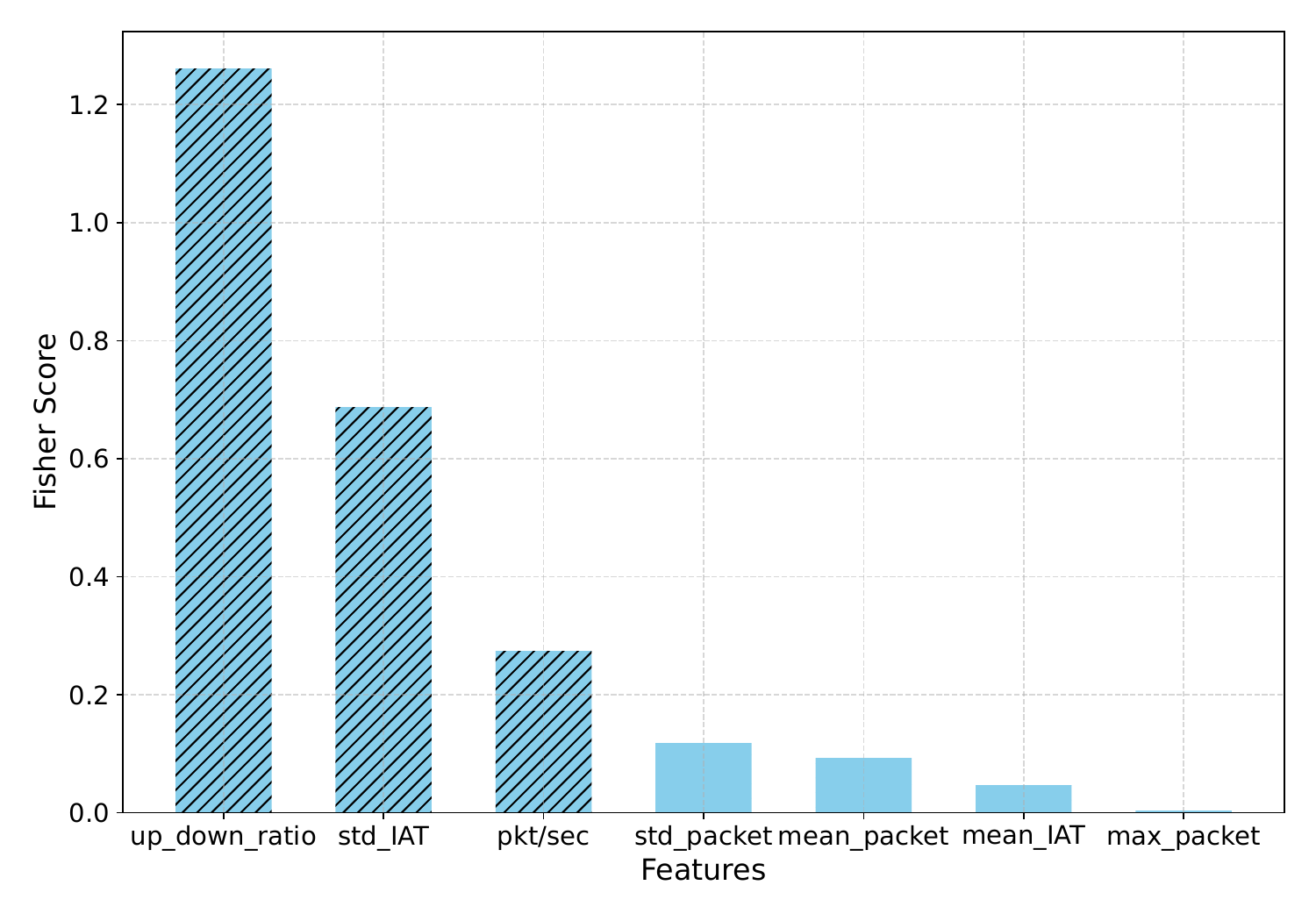}}\label{fig:fshr}
\hspace{-0.1 in}
\hfill
\subfigure[]
{\includegraphics[width=0.24\textwidth]{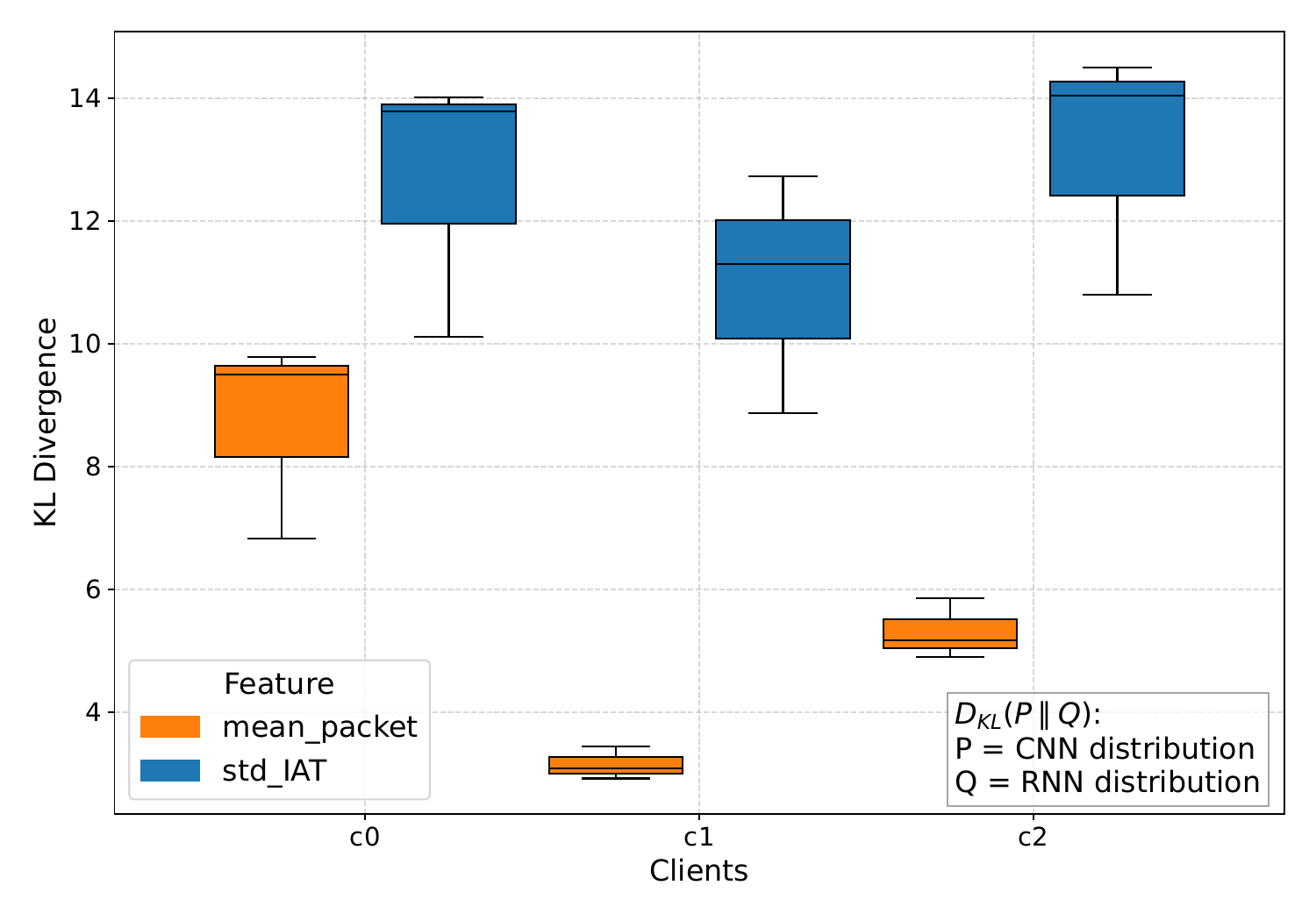}}\label{fig:kldv}
\vspace{-0.12 in}
\caption{Feature separability analysis. 
    (a) Fisher Score plot, showing the importance of different features
    (b) KL divergence analysis between CNN and RNN. (c0 indicates Jetson Orin, c1 is M1 chip-based MacBook, and c2 is an Intel-based laptop).}
\label{fig:fetrs_sep}   
\end{figure}

Here, we quantitatively and visually analyze the extracted traffic features to assess the separability of CNN- and RNN-based architectures. We utilize the Fisher score to determine the most essential features that are necessary for the fingerprinting task. In~\ref{fig:fetrs_sep}(a), we present a few critical features, which have relatively high Fisher score values.

To further quantify feature-level divergence between CNN and RNN traffic, we compute the \textit{Kullback--Leibler (KL) divergence} between distributions of key statistical features, including \textit{mean packet size}, \textit{mean interarrival time}, and \textit{standard deviation of interarrival time}. Let \( P(x) \) and \( Q(x) \) denote the empirical distributions of a given feature extracted from traffic traces of two different model classes. Therefore, the KL divergence is defined as:
\vspace{-0.05in}
\begin{equation} \label{eu_eqn}
D_{\text{KL}}(P \parallel Q) = \sum_{x} P(x) \log \frac{P(x)}{Q(x)}.
\end{equation}
\vspace{-0.12in}

A higher KL divergence indicates greater distinguishability between the two distributions, suggesting that the corresponding feature captures architecture-specific traffic patterns. To preserve model-level variability, we compute average KL divergence \textit{per trace} using equation~\ref{eu_eqn}, comparing each trace from one architecture class against all traces from the opposing class. This results in a distribution of KL scores per client, visualized as box plots in Fig.~\ref{fig:fetrs_sep}(b).

\begin{table}[!t]
\caption{Mean $\pm$ std of KL divergence scores per client and feature.}
\emph{Note:} “Model (P)” is the reference distribution \(P\); the comparison distribution \(Q\) is the other architecture (e.g., P=CNN, Q=RNN).
\centering
\label{tab:my-KLT}
\resizebox{1\columnwidth}{!}{
\begin{tabular}{ccccc}
\hline
\multirow{2}{*}{Client} & \multirow{2}{*}{Reference Model (P)} & \multicolumn{3}{c}{KL Divergence} \\ \cline{3-5}
 &  & Mean Frame & Mean IAT & Std IAT \\ \hline
\multirow{2}{*}{C0 (Orin)} & CNN & 8.70 ± 1.63 & 12.73 ± 0.60 & 12.64 ± 2.19 \\ \cline{2-5}
 & RNN & 0.49 ± 0.14 & 2.75 ± 3.80 & 5.30 ± 3.91 \\ \hline
\multirow{2}{*}{C1 (MacBook)} & CNN & 3.15 ± 0.27 & 1.44 ± 0.31 & 10.97 ± 1.95 \\ \cline{2-5}
 & RNN & 1.07 ± 0.53 & 3.17 ± 3.92 & 5.63 ± 4.09 \\ \hline
\multirow{2}{*}{C2 (Laptop)} & CNN & 5.31 ± 0.49 & 5.74 ± 2.36 & 13.11 ± 2.02 \\ \cline{2-5}
 & RNN & 0.89 ± 0.52 & 1.83 ± 2.53 & 9.20 ± 7.32 \\ \hline
\end{tabular}
}\vspace{-0.1in}
\end{table}

This demonstrates that clients consistently exhibit higher KL divergence values when the reference distribution is from CNN traces. In particular, the Jetson Orin client (c0) shows the most distinguishable patterns, with mean KL values exceeding 8.7 and 12.7 for packet size and IAT, respectively.  These trends are quantitatively summarized in Table~\ref{tab:my-KLT}, which reports the mean and standard deviation of KL divergence for each client and feature. This highlights that certain statistical features, especially those capturing timing dynamics and payload volume, exhibit strong discriminative behavior between CNN and RNN models. Lower KL values for RNN traces suggest that their traffic is comparatively more uniform or less bursty, making them less distinguishable on these features alone.
\vspace{-0.05in}

\begin{figure}[h]
\centering
\subfigure[]
{\includegraphics[width=0.24\textwidth]{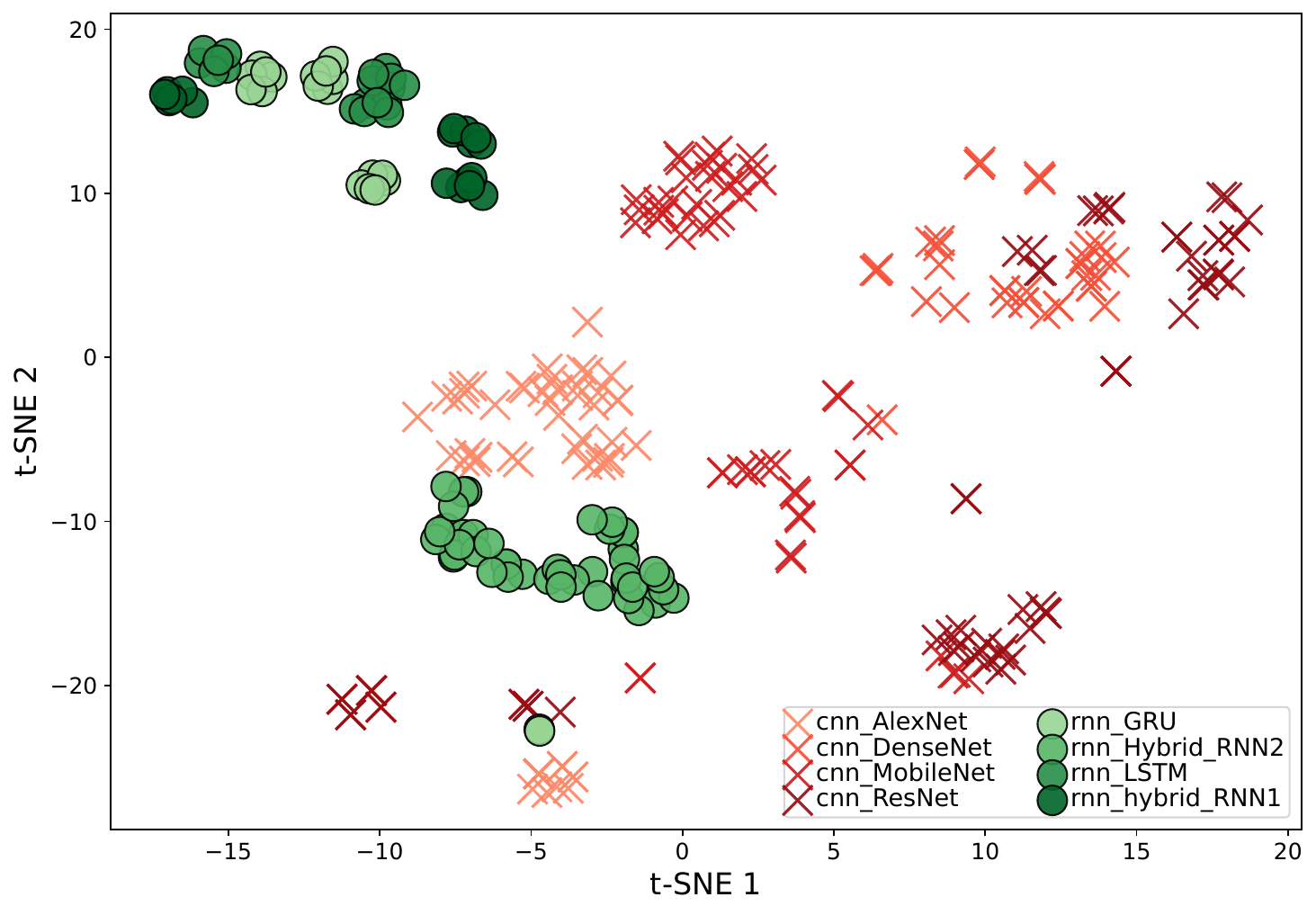}\label{fig:sktrplt}}
\hspace{-0.1 in}
\hfill
\subfigure[]
{\includegraphics[width=0.24\textwidth]{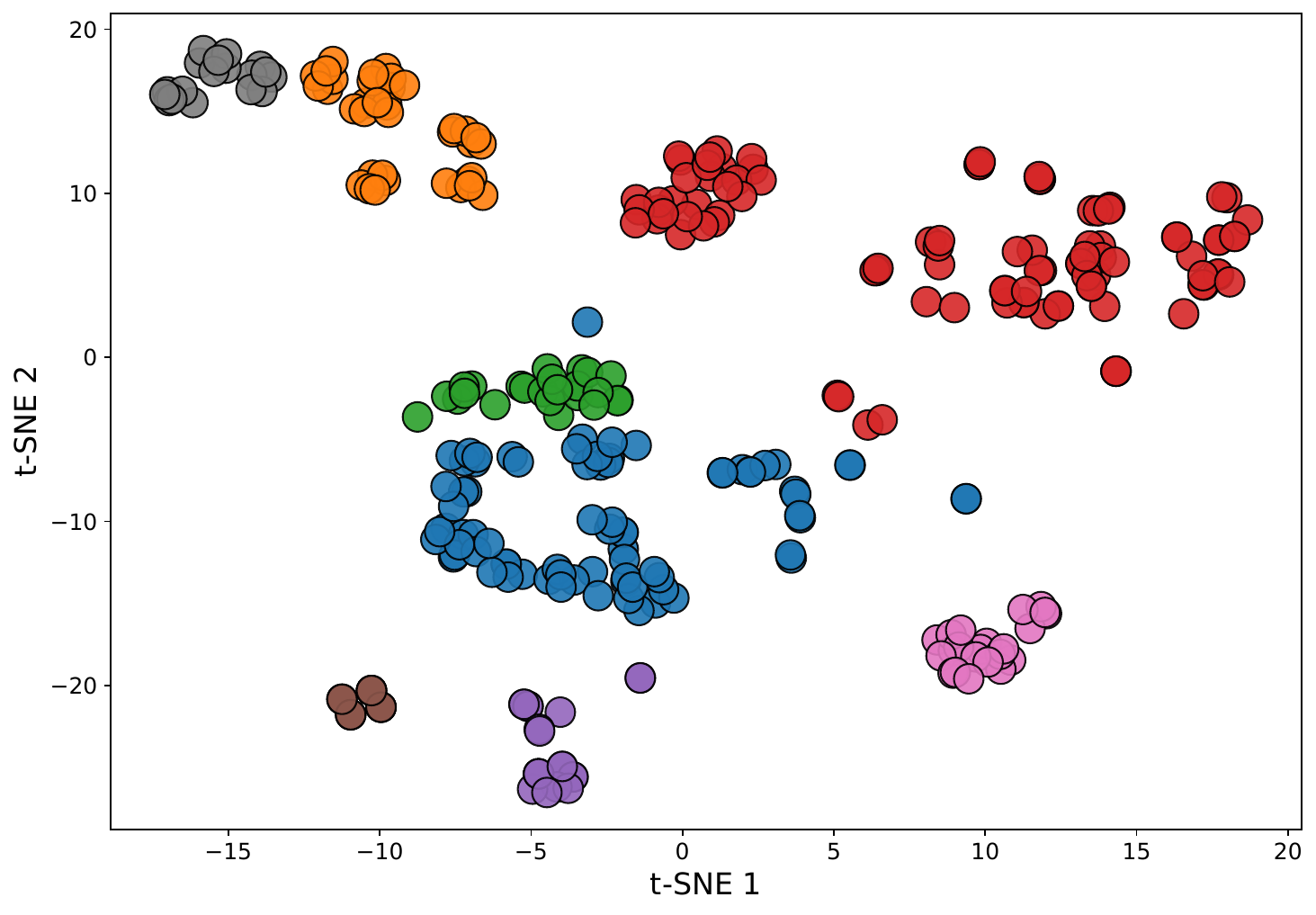}\label{fig:clstplt}}
\vspace{-0.12 in}
\caption{Feature visualization and clustering analysis of CNN vs. RNN architectures. 
    (a) Two-dimensional t-SNE embedding of extracted network traffic features, demonstrating clear visual separability between CNN- and RNN-based architectures.
    (b) Corresponding spectral clustering results ($k=8$), showing distinct clusters aligned closely with CNN and RNN classes.}
\label{fig:fetr-plot}   
\end{figure}

\vspace{-0.05in}
In addition to KL analysis, we apply t-distributed Stochastic Neighbor Embedding (t-SNE) to visualize the non-linear structure in the extracted features. As shown in Fig.~\ref {fig:fetr-plot}(a), t-SNE reveals distinct clusters corresponding to CNN- and RNN-based models. Further validation using K-means clustering (Fig.~\ref {fig:fetr-plot}(b), k=8) supports this observation, showing two dominant clusters aligned with the respective architecture families. However, fine-grained intra-class distinctions (e.g., ResNet vs. DenseNet) remain less pronounced. During the observation period, it is also possible for some traces to resemble the patterns of CNNs—particularly because interarrival time has a strong influence on model classification. As a result, in Fig.~\ref {fig:fetr-plot}, some RNN traces appear very close to CNN clusters. Together, these statistical and visual results support the feasibility of fingerprinting broad architectural modalities using network traffic features.

\textbf{Insight:} Through Fisher score analysis, we observe that features such as inter-arrival time and packet size consistently differ between CNN and RNN traffic patterns. These differences are supported both statistically and visually—via KL divergence, clustering, and t-SNE projections. The consistency of these patterns across clients confirms that deep learning architectures influence communication behavior in ways that can be identified through passive traffic analysis, even without access to packet contents.

\subsection{Effect of Fingerprinting on Resource Denial Attack} 
Resource denial attacks are emulated on our testbed by throttling the effective throughput of both CNN and RNN-based FL applications, thereby causing a delay in global convergence. Convergence is reached once the global model attains $90\%$ accuracy. Experiments show that the uncompromised CNN and RNN models require $216\pm192$~Mbps and $384\pm56$~Mbps of mean effective throughput. CNNs generally require significantly more training rounds than RNNs; moreover, their packet sizes vary substantially because of fragmentation. Therefore, we see a higher standard deviation in the mean effective throughput requirement for CNN. 

\begin{figure}[htb!]
\centering
\subfigure[]
{\includegraphics[width=0.235\textwidth]{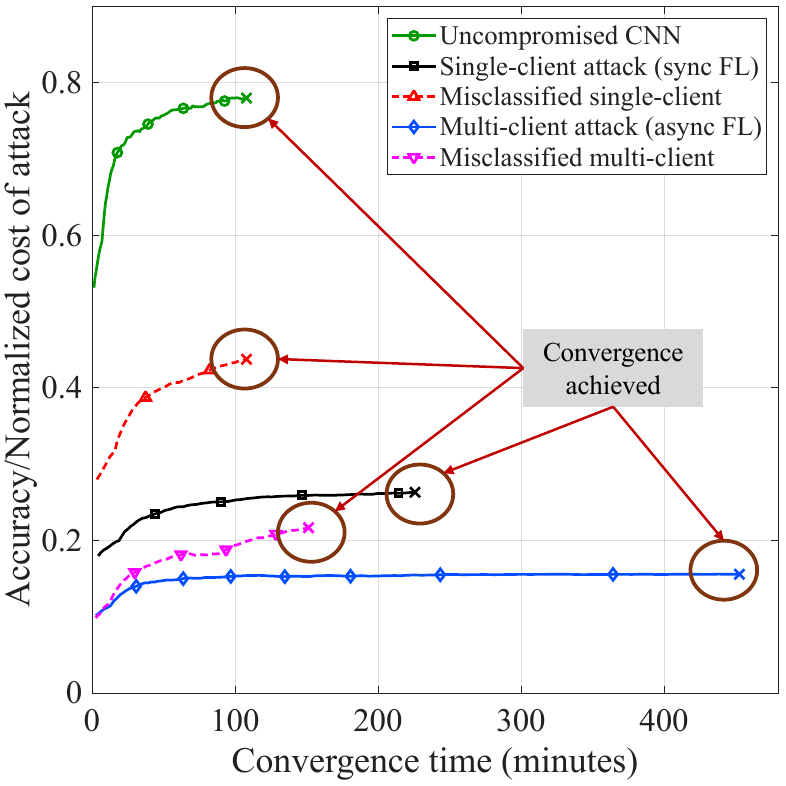}\label{fig:cnnAttack}}
\hspace*{\fill}
\subfigure[]
{\includegraphics[width=0.24\textwidth]{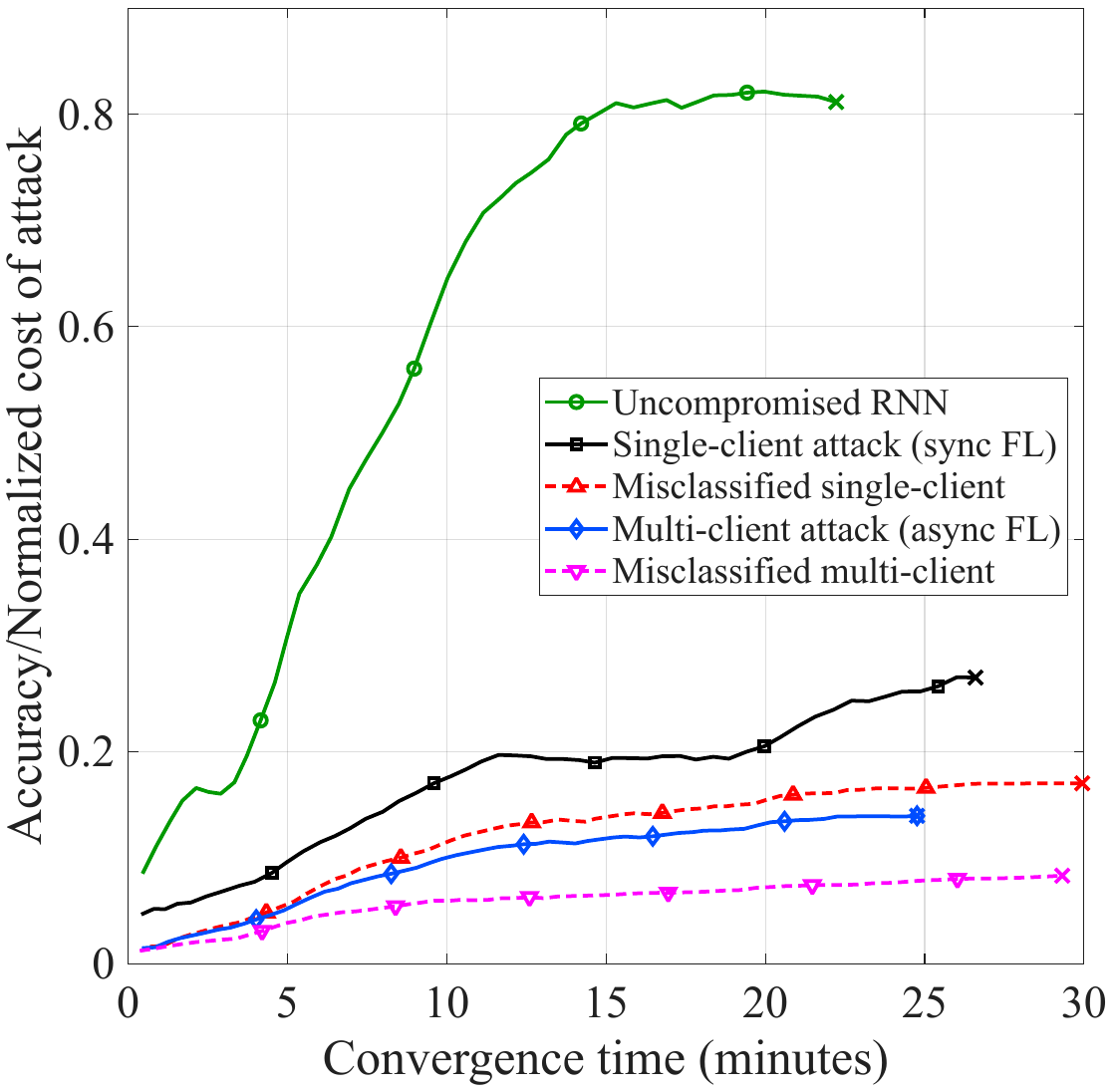}\label{fig:rnnAttack}}
\vspace{-0.12 in}
\caption{Resource denial attack on (a) CNN and (b) RNN-based FL applications after successful and unsuccessful fingerprinting.}
\label{fig:CNNattackRNN}   
\end{figure}

For this analysis, we consider both synchronous and asynchronous global model updates. In synchronous mode, the server waits for local model updates from all the clients before aggregating and publishing the next global model. Whereas, in asynchronous mode, the server updates its global model as soon as it receives any update from any of the participating clients. The attacks are devised in our experiment in a way that a single or multiple clients are deprived of two-thirds of the required throughput. Fig. \ref{fig:CNNattackRNN} shows the convergence time of all the models in terms of accuracy/normalized cost of attack. The accuracy value is taken as a 15-point moving-averaged mean. In addition, the cost of attack is $c_A = n_c\times\frac{B_{uncomp}}{B_{denied}}$, where $B_{uncomp}$ and $B_{denied}$ are the mean effective throughput of uncompromised and attacked clients, respectively, and the number of clients under attack is $n_c$. The normalized cost for uncompromised applications is $1$.
\vspace{-0.1in}

In attacks enabled with successful fingerprinting, a considerable amount of convergence delay is observed. For instance, single (sync FL) and multi-client (async FL) attacks on CNN models increase the convergence delay by $109\%$ and $319\%$, respectively, and attacks on RNN models increase the delay by $20\%$ and $19\%$. This large difference between the increase in delays in CNN and RNN models is mainly because CNN models generally have larger packets, which are more vulnerable to resource denial attacks. Additionally, there are two scenarios for each model where the applications are misclassified (i.e., CNN as RNN and vice-versa), leading to unsuccessful and ineffective attacks.

\textbf{Insight:} When a CNN model is attacked after being fingerprinted as an RNN, the convergence delay increases by $0.1\%$ and $40\%$ for single and multi-client attacks, which are far less than what a successfully fingerprinted attack would experience. However, when an RNN is misclassified, the convergence delay indeed increases more, but the cost is around $167\%$ higher than what a correctly classified attack requires. Therefore, it is instrumental to fingerprint models with high accuracy for an effective resource denial attack.


\vspace{-0.05in}

\section{Discussion}

Our findings demonstrate that Deep Learning architectures in FL—specifically CNNs and RNNs—produce statistically distinct communication patterns that can be exploited through passive traffic analysis. Since architecture often correlates with data modality (e.g., images vs. sequences), identifying the model also reveals indirect information about the nature of client data. This dual leakage poses a significant privacy threat.
\vspace{-0.15in}

\subsection{Attack Implications}
FLARE’s ability to infer architecture passively enables more precise downstream attacks. Adversaries can craft adversarial examples using architecture-aligned surrogate models, or apply inversion and membership inference techniques optimized for CNNs or RNNs. These targeted attacks are more effective than generic black-box strategies, and the passive nature of FLARE makes detection difficult. Moreover, because architecture is often linked to application context, fingerprinting may also compromise client intent or behavior.
\vspace{-0.05in}

\subsection{Countermeasures}
As shown earlier, FLARE remains effective in both open- and closed-world settings. Mitigating such a robust fingerprinting attack requires defenses that disrupt traffic-level leakage without significantly degrading FL performance. Packet padding and randomized update schedules can obscure temporal and size-based patterns~\cite{wright2008spot}, but often increase bandwidth and latency—undesirable for edge devices. Secure aggregation protocols~\cite{bonawitz2017practical} reduce visibility into individual updates but introduce synchronization overhead. Sparsification and quantization~\cite{pokhrel2020federated} can reduce traffic diversity and fingerprintability, though they may degrade FL performance under non-IID data. MAC address randomization helps anonymize clients but is not uniformly supported across edge devices. These limitations highlight the need for lightweight, practical defenses that preserve privacy while maintaining efficiency in FL systems.
\vspace{-0.05in}

\section{Conclusion}
We proposed FLARE, a fingerprinting framework that identifies deep learning architectures in federated learning by analyzing network traffic patterns. Our approach combines flow-level and packet-level features to capture detailed traffic characteristics and uses late fusion meta-models to enhance attack performance. Through extensive experiments in both close-world and open-world settings, FLARE demonstrates strong generalization through high precision, recall, and F1 scores. Incorporating device heterogeneity during training further improves system robustness. This work is the first to uncover a critical and previously unexplored privacy risk in wireless federated learning, which is the ability to identify deep learning model architectures solely through passive observation of network traffic. Our findings underscore the need for either application or network-level defenses against side-channel leakage in FL systems.


\bibliographystyle{IEEEtran}
\bibliography{reference}

@inproceedings{mcmahan2017communication,
  title={Communication-efficient learning of deep networks from decentralized data},
  author={McMahan, Brendan and Moore, Eider and Ramage, Daniel and Hampson, Seth and y Arcas, Blaise Aguera},
  booktitle={Proc. Artificial intelligence and statistics},
  pages={1273--1282},
  year={2017},
  organization={PMLR}
}

@article{oh2021gandalf,
  title={GANDaLF: GAN for data-limited fingerprinting},
  author={Oh, Se Eun and Mathews, Nate and Rahman, Mohammad Saidur and Wright, Matthew and Hopper, Nicholas},
  journal={Proceedings on privacy enhancing technologies},
  volume={2021},
  number={2},
  year={2021}
}

@article{kairouz2021advances,
  title={Advances and open problems in federated learning},
  author={Kairouz, Peter and McMahan, H Brendan and Avent, Brendan and Bellet, Aur{\'e}lien and Bennis, Mehdi and Bhagoji, Arjun Nitin and Bonawitz, Kallista and Charles, Zachary and Cormode, Graham and Cummings, Rachel and others},
  journal={Foundations and trends{\textregistered} in machine learning},
  volume={14},
  number={1--2},
  pages={1--210},
  year={2021},
  publisher={Now Publishers, Inc.}
}

@article{kaushal2025securing,
  title={Securing the collective intelligence: a comprehensive review of federated learning security attacks and defensive strategies},
  author={Kaushal, Vishal and Sharma, Sangeeta},
  journal={Knowledge and Information Systems},
  pages={1--39},
  year={2025},
  publisher={Springer}
}

@article{zhang2024survey,
  title={A survey of trustworthy federated learning: Issues, solutions, and challenges},
  author={Zhang, Yifei and Zeng, Dun and Luo, Jinglong and Fu, Xinyu and Chen, Guanzhong and Xu, Zenglin and King, Irwin},
  journal={ACM Transactions on Intelligent Systems and Technology},
  volume={15},
  number={6},
  pages={1--47},
  year={2024},
  publisher={ACM New York, NY, USA}
}

@article{zhang2025physical,
  title={Physical Layer-Based Device Fingerprinting For Wireless Security: From Theory To Practice},
  author={Zhang, Junqing and Ardizzon, Francesco and Piana, Mattia and Shen, Guanxiong and Tomasin, Stefano},
  journal={IEEE Transactions on Information Forensics and Security},
  year={2025},
  publisher={IEEE}
}

@article{sheng2025network,
  title={Network traffic fingerprinting for IIoT device identification: A survey},
  author={Sheng, Chuan and Zhou, Wei and Han, Qing-Long and Ma, Wanlun and Zhu, Xiaogang and Wen, Sheng and Xiang, Yang},
  journal={IEEE Transactions on Industrial Informatics},
  year={2025},
  publisher={IEEE}
}

@inproceedings{acar2020peek,
  title={Peek-a-boo: I see your smart home activities, even encrypted!},
  author={Acar, Abbas and Fereidooni, Hossein and Abera, Tigist and Sikder, Amit Kumar and Miettinen, Markus and Aksu, Hidayet and Conti, Mauro and Sadeghi, Ahmad-Reza and Uluagac, Selcuk},
  booktitle={Proc. the 13th ACM Conference on Security and Privacy in Wireless and Mobile Networks},
  pages={207--218},
  year={2020}
}

@inproceedings{wang2020high,
  title={High precision open-world website fingerprinting},
  author={Wang, Tao},
  booktitle={Proc. 2020 IEEE Symposium on Security and Privacy (SP)},
  pages={152--167},
  year={2020},
  organization={IEEE}
}

@inproceedings{panchenko2016website,
  title={Website Fingerprinting at Internet Scale.},
  author={Panchenko, Andriy and Lanze, Fabian and Pennekamp, Jan and Engel, Thomas and Zinnen, Andreas and Henze, Martin and Wehrle, Klaus},
  booktitle={Proc. NDSS},
  volume={1},
  pages={23477},
  year={2016}
}

@article{laperdrix2020browser,
  title={Browser fingerprinting: A survey},
  author={Laperdrix, Pierre and Bielova, Nataliia and Baudry, Benoit and Avoine, Gildas},
  journal={ACM Transactions on the Web (TWEB)},
  volume={14},
  number={2},
  pages={1--33},
  year={2020},
  publisher={ACM New York, NY, USA}
}

@article{nayak2020data,
  title={Data leakage detection and prevention: Review and research directions},
  author={Nayak, Suvendu Kumar and Ojha, Ananta Charan},
  journal={Machine learning and information processing: proceedings of ICMLIP 2019},
  pages={203--212},
  year={2020},
  publisher={Springer}
}

@article{yang2025deep,
  title={Deep learning model inversion attacks and defenses: a comprehensive survey},
  author={Yang, Wencheng and Wang, Song and Wu, Di and Cai, Taotao and Zhu, Yanming and Wei, Shicheng and Zhang, Yiying and Yang, Xu and Tang, Zhaohui and Li, Yan},
  journal={Artificial Intelligence Review},
  volume={58},
  number={8},
  pages={1--52},
  year={2025},
  publisher={Springer}
}

@article{breiman2001random,
  title={Random forests},
  author={Breiman, Leo},
  journal={Machine learning},
  volume={45},
  pages={5--32},
  year={2001},
  publisher={Springer}
}

@inproceedings{ngiam2011multimodal,
  title={Multimodal deep learning.},
  author={Ngiam, Jiquan and Khosla, Aditya and Kim, Mingyu and Nam, Juhan and Lee, Honglak and Ng, Andrew Y and others},
  booktitle={Proc. ICML},
  volume={11},
  pages={689--696},
  year={2011}
}

@article{baltruvsaitis2018multimodal,
  title={Multimodal machine learning: A survey and taxonomy},
  author={Baltru{\v{s}}aitis, Tadas and Ahuja, Chaitanya and Morency, Louis-Philippe},
  journal={IEEE Transactions on Pattern Analysis and Machine Intelligence},
  volume={41},
  number={2},
  pages={423--443},
  year={2018},
  publisher={IEEE}
}

@article{ramachandram2017deep,
  title={Deep multimodal learning: A survey on recent advances and trends},
  author={Ramachandram, Dhanesh and Taylor, Graham W},
  journal={IEEE signal processing magazine},
  volume={34},
  number={6},
  pages={96--108},
  year={2017},
  publisher={IEEE}
}

@inproceedings{wright2008spot,
  title={Spot me if you can: Uncovering spoken phrases in encrypted voip conversations},
  author={Wright, Charles V and Ballard, Lucas and Coull, Scott E and Monrose, Fabian and Masson, Gerald M},
  booktitle={Proc. 2008 IEEE Symposium on Security and Privacy (sp 2008)},
  pages={35--49},
  year={2008},
  organization={IEEE}
}

@inproceedings{bonawitz2017practical,
  title={Practical secure aggregation for privacy-preserving machine learning},
  author={Bonawitz, Keith and Ivanov, Vladimir and Kreuter, Ben and Marcedone, Antonio and McMahan, H Brendan and Patel, Sarvar and Ramage, Daniel and Segal, Aaron and Seth, Karn},
  booktitle={Proc. the 2017 ACM SIGSAC Conference on Computer and Communications Security},
  pages={1175--1191},
  year={2017}
}

@inproceedings{nasr2019comprehensive,
  title={Comprehensive privacy analysis of deep learning: Passive and active white-box inference attacks against centralized and federated learning},
  author={Nasr, Milad and Shokri, Reza and Houmansadr, Amir},
  booktitle={Proc. 2019 IEEE symposium on security and privacy (SP)},
  pages={739--753},
  year={2019},
  organization={IEEE}
}

@article{pokhrel2020federated,
  title={Federated learning with blockchain for autonomous vehicles: Analysis and design challenges},
  author={Pokhrel, Shiva Raj and Choi, Jinho},
  journal={IEEE Transactions on Communications},
  volume={68},
  number={8},
  pages={4734--4746},
  year={2020},
  publisher={IEEE}
}

@article{pillutla2022robust,
  title={Robust aggregation for federated learning},
  author={Pillutla, Krishna and Kakade, Sham M and Harchaoui, Zaid},
  journal={IEEE Transactions on Signal Processing},
  volume={70},
  pages={1142--1154},
  year={2022},
  publisher={IEEE}
}

@article{yuan2022convergence,
  title={On convergence of fedprox: Local dissimilarity invariant bounds, non-smoothness and beyond},
  author={Yuan, Xiaotong and Li, Ping},
  journal={Advances in Neural Information Processing Systems},
  volume={35},
  pages={10752--10765},
  year={2022}
}

@inproceedings{mehnaz2022your,
  title={Are your sensitive attributes private? novel model inversion attribute inference attacks on classification models},
  author={Mehnaz, Shagufta and Dibbo, Sayanton V and Kabir, Ehsanul and Li, Ninghui and Bertino, Elisa},
  booktitle={Proc. 31st USENIX Security Symposium (USENIX Security 22)},
  pages={4579--4596},
  year={2022}
}

@article{wang2022poisoning,
  title={Poisoning-assisted property inference attack against federated learning},
  author={Wang, Zhibo and Huang, Yuting and Song, Mengkai and Wu, Libing and Xue, Feng and Ren, Kui},
  journal={IEEE Transactions on Dependable and Secure Computing},
  volume={20},
  number={4},
  pages={3328--3340},
  year={2022},
  publisher={IEEE}
}

@inproceedings{zhu2025fedmia,
  title={FedMIA: An Effective Membership Inference Attack Exploiting" All for One" Principle in Federated Learning},
  author={Zhu, Gongxi and Li, Donghao and Gu, Hanlin and Yao, Yuan and Fan, Lixin and Han, Yuxing},
  booktitle={Proc. the Computer Vision and Pattern Recognition Conference},
  pages={20643--20653},
  year={2025}
}

@inproceedings{chen2025deep,
  title={Deep Diffusion Gradients Leakage in Federated Learning},
  author={Chen, Dexuan and Luo, Yueyi and Qi, Qianqian and Fei, Hongxiao},
  booktitle={Proc. ICASSP 2025-2025 IEEE International Conference on Acoustics, Speech and Signal Processing (ICASSP)},
  pages={1--5},
  year={2025},
  organization={IEEE}
}

@inproceedings{shokri2017membership,
  title={Membership inference attacks against machine learning models},
  author={Shokri, Reza and Stronati, Marco and Song, Congzheng and Shmatikov, Vitaly},
  booktitle={Proc. 2017 IEEE Symposium on Security and Privacy (SP)},
  pages={3--18},
  year={2017},
  organization={IEEE}
}

@inproceedings{batina2019csi,
  title={$\{$CSI$\}$$\{$NN$\}$: Reverse engineering of neural network architectures through electromagnetic side channel},
  author={Batina, Lejla and Bhasin, Shivam and Jap, Dirmanto and Picek, Stjepan},
  booktitle={Proc. 28th USENIX Security Symposium (USENIX Security 19)},
  pages={515--532},
  year={2019}
}

@inproceedings{zhang2023deep,
  title={Deep-learning model extraction through software-based power side-channel},
  author={Zhang, Xiang and Ding, Aidong Adam and Fei, Yunsi},
  booktitle={Proc. 2023 IEEE/ACM International Conference on Computer Aided Design (ICCAD)},
  pages={1--9},
  year={2023},
  organization={IEEE}
}

@article{chabanne2021side,
  title={Side channel attacks for architecture extraction of neural networks},
  author={Chabanne, Herv{\'e} and Danger, Jean-Luc and Guiga, Linda and K{\"u}hne, Ulrich},
  journal={CAAI Transactions on Intelligence Technology},
  volume={6},
  number={1},
  pages={3--16},
  year={2021},
  publisher={Wiley Online Library}
}

@article{li2024robust,
  title={Robust App Fingerprinting Over the Air},
  author={Li, Jianfeng and Lin, Zheng and Qu, Jian and Wu, Shuohan and Zhou, Hao and Liu, Yangyang and Ma, Xiaobo and Wang, Ting and Luo, Xiapu and Guan, Xiaohong},
  journal={IEEE/ACM Transactions on Networking},
  year={2024},
  publisher={IEEE}
}

@article{chowdhury2022survey,
  title={A survey on device fingerprinting approach for resource-constraint IoT devices: Comparative study and research challenges},
  author={Chowdhury, Rajarshi Roy and Abas, Pg Emeroylariffion},
  journal={Internet of Things},
  volume={20},
  pages={100632},
  year={2022},
  publisher={Elsevier}
}

@article{wang2019edge,
  title={In-edge ai: Intelligentizing mobile edge computing, caching and communication by federated learning},
  author={Wang, Xiaofei and Han, Yiwen and Wang, Chenyang and Zhao, Qiyang and Chen, Xu and Chen, Min},
  journal={Ieee Network},
  volume={33},
  number={5},
  pages={156--165},
  year={2019},
  publisher={IEEE}
}

@inproceedings{sirinam2019triplet,
  title={Triplet fingerprinting: More practical and portable website fingerprinting with n-shot learning},
  author={Sirinam, Payap and Mathews, Nate and Rahman, Mohammad Saidur and Wright, Matthew},
  booktitle={Proceedings of the 2019 ACM SIGSAC Conference on Computer and Communications Security},
  pages={1131--1148},
  year={2019}
}

\end{document}